\documentclass[reprint,twocolumn,preprintnumbers,amsmath,amssymb,superscriptaddress,aps,a4paper,prb]{revtex4}

\newcommand{\iga}[2]{In\ensuremath{_{#1}}Ga\ensuremath{_{#2}}As}
\newcommand{\iaa}[2]{In\ensuremath{_{#1}}Al\ensuremath{_{#2}}As}
\newcommand{\pedi}[1]{\ensuremath{_{#1}}}
\newcommand{\apic}[1]{\ensuremath{^{#1}}}
\renewcommand{\eqref}[1]{Eq.\ \ref{#1}}

\usepackage[dvips]{color}
\usepackage{graphicx}% Include figure files
\usepackage{amsmath}
\usepackage{amssymb}
\usepackage{amsfonts}
\usepackage{amsthm}
\usepackage{lmodern}
\usepackage{textcomp}
\usepackage[T1]{fontenc}
\usepackage{gensymb} %use \micro, \ohm, \degree, \celsius!
\usepackage[seperr, load={}]{siunitx}

\begin{document}

\title{Quantum dot spectroscopy of proximity-induced superconductivity in a two-dimensional electron gas}

\author{F. Deon}
	\email{f.deon@sns.it}
	\affiliation{NEST, Istituto Nanoscienze-CNR  and Scuola Normale Superiore, I-56127 Pisa, Italy}
\author{V. Pellegrini}
	\affiliation{NEST, Istituto Nanoscienze-CNR  and Scuola Normale Superiore, I-56127 Pisa, Italy}\date{\today}
\author{F. Giazotto}
	\affiliation{NEST, Istituto Nanoscienze-CNR  and Scuola Normale Superiore, I-56127 Pisa, Italy}\date{\today}
\author{G. Biasiol} 
		\affiliation{CNR-IOM, Laboratorio TASC, Area Science Park, I-34149 Trieste, Italy}
\author{L. Sorba}
	\affiliation{NEST, Istituto Nanoscienze-CNR  and Scuola Normale Superiore, I-56127 Pisa, Italy}
\author{F. Beltram}
	\affiliation{NEST, Istituto Nanoscienze-CNR  and Scuola Normale Superiore, I-56127 Pisa, Italy}
%\pacs{73.23.Hk}% insert suggested PACS numbers in braces on next line

\date{\today}      

\begin{abstract}
We report the realization of a hybrid superconductor-quantum dot device by means of top-down nanofabrication starting from a two dimensional electron gas in a InGaAs/InAlAs semiconductor heterostructure. The quantum dot is defined by electrostatic gates placed within the normal region of a planar Nb-InGaAs quantum well-Nb junction. Measurements in the regime of strong Coulomb blockade as well as cotunneling spectroscopy allow to directly probe the proximity-induced energy gap in a ballistic two-dimensional electron gas coupled to superconductors.
\end{abstract}

\maketitle

Hybrid devices in which a quantum dot (QD) is connected to superconducting (S) electrodes display a rich physical behavior, which stems from the coexistence of competing phenomena such as proximity superconductivity, Coulomb blockade and the Kondo effect\cite{ref1}. Several examples of hybrid QD devices have been demonstrated so far, all of which rely on nanostructures defined by bottom-up approaches such as carbon nanotubes\cite{ref2}, semiconductor nanowires\cite{ref3} and self-assembled QDs\cite{ref4,ref5}. Transport studies in these nanosystems have demonstrated Josephson currents in S-QD-S devices and tuning of the critical current by changing the QD charge state\cite{ref6,ref7}. Other authors have focused on quasiparticle transport, in particular on the interplay of Coulomb blockade, superconducting order in the leads and Kondo effect\cite{ref8,ref9}. 

In the perspective of practical implementations, large-scale integrability is a key requirement, and a top-down nanofabrication approach based on two dimensional electron gases (2DEGs) confined in semiconductor heterostructures would be preferable. Finally, QDs embedded into 2DEG-based hybrid structures can be used as a spectroscopic tool of proximity-induced superconductivity in ballistic electronic systems. While the local density of states (DOS) in proximized normal metal films has been probed using weakly-coupled tunnel contacts\cite{ref10,ref11}, a similar measurement in the case of the 2DEG hosted in a ballistic semiconductor heterostructure has not been performed to date.

In this letter we report the realization of a hybrid S-QD-S device by means of top-down nanofabrication, starting from a 2DEG confined in a \iga{0.80}{0.20}/\iaa{0.75}{0.25} heterostructure. Following previous work\cite{ref12,ref13,ref14}, the QD is formed by applying negative voltages to surface electrostatic gates placed on top of a narrow mesa strip etched in the heterostructure, and laterally contacted by superconducting niobium electrodes. We discuss the impact of the proximity-induced superconductivity on the transport properties in different regimes of coupling to the leads, and measure the proximity-induced energy gaps in the two-dimensional electron gas.

A scanning electron micrograph and the main nanofabrication steps are shown in Fig. 1(a,b). The active region of the heterostructure contains a \SI{15}{nm}-thick $\delta$-doped \iga{0.80}{0.20} quantum well\cite{ref15,ref16}, sandwiched  between \iaa{0.75}{0.25} barriers. The sheet electron density, measured from the period of Shubnikov - de Haas oscillations, is $n_S\simeq\SI{5.9E+11}{cm^{-2}}$. The mobility $\mu\simeq$\SI{1.8E+5}{cm^2V^{-1}s^{-1}} is equivalent to an electron mean free path $l_p\simeq\SI{2.3}{\micro m}$. InAs and \iga{x}{1-x} alloys with high molar fraction $x\geq0.75$ have often been used in association with Nb for the realization of hybrid ballistic devices, thanks to their property of forming Schottky barrier-free junctions. For the same reason a suitable dielectric is required for the electrical insulation of the electrostatic gates.

The nanofabrication of the hybrid device requires several mutually-aligned steps of electron beam lithography (EBL). First [Fig. 1(a1)] we pattern a dielectric strip on the heterostructure surface, by means of EBL, using hydrogen silsesquioxane (HSQ) as a negative tone e-beam resist\cite{ref17}. Thickness and width of the strip are \SI{60}{nm} and \SI{600}{nm}, respectively. The mesa strip is defined by wet etching in a H\pedi{3}PO\pedi{4}-H\pedi{2}O\pedi{2} solution through an EBL-patterned poly(methyl methacrylate) mask (a2). Electrostatic gates are then added by thermal evaporation and liftoff (a3), followed by sputtering and liftoff of the Nb side contacts (a4). The junction has a width $W=\SI{3.0}{\micro m}$, while the inter-electrode distance, equal to the width of the etched mesa strip, is $L=\SI{650}{nm}$. In order to maximize interface transmissivity, the native oxide and possible contaminants must be removed from the mesa sidewalls before the deposition of the contacts. To this end we perform low-energy Ar\apic{+} sputter-cleaning in high vacuum, immediately  before the deposition of Nb\cite{ref18}$^,$ \cite{ref18.1}.

Low temperature measurements are performed down to \SI{150}{mK}. Below the critical temperature $T_c\approx 9K$ of the Nb films, proximity-induced superconductivity\cite{ref18.5} in the semiconductor modifies the DOS at energies $|\epsilon|\lesssim\Delta$ (where $\Delta= 1.764k_BT_c\simeq\SI{1.35}{meV}$ is the superconducting gap of the Nb films and $k_B$ the Boltzmann constant).  The normal-state resistance of the junction is $R_N=$\SI{150}{\ohm}, roughly two times larger than the lower bound set by the Sharvin resistance $R_{Sh}=(\pi h)/(2e^2W\sqrt{2\pi n_S})=$\SI{70}{\ohm}, where $h$ is Planck's constant. The voltage-current characteristic\cite{ref18.6} of the open junction (all gates set to ground), measured at \SI{235}{mK}, is plotted in Fig. 1(c) and shows a switching current $I_c\simeq\SI{800}{nA}$. We note that in this configuration the junction length $L$ is smaller than the electron scattering length $l_p$ and larger than the induced coherence length $\xi_N=(\hbar^2 \sqrt{2\pi n_S})/(\Delta m^*)=\SI{360}{nm}$ (where $m^*=0.03 m_e$ is the electron effective mass in \iga{0.80}{0.20}), so that the normal region is ballistic and the junction falls in the intermediate length regime. 

\begin{figure}[t!]
\label{Fig1}
\includegraphics[width=85mm]{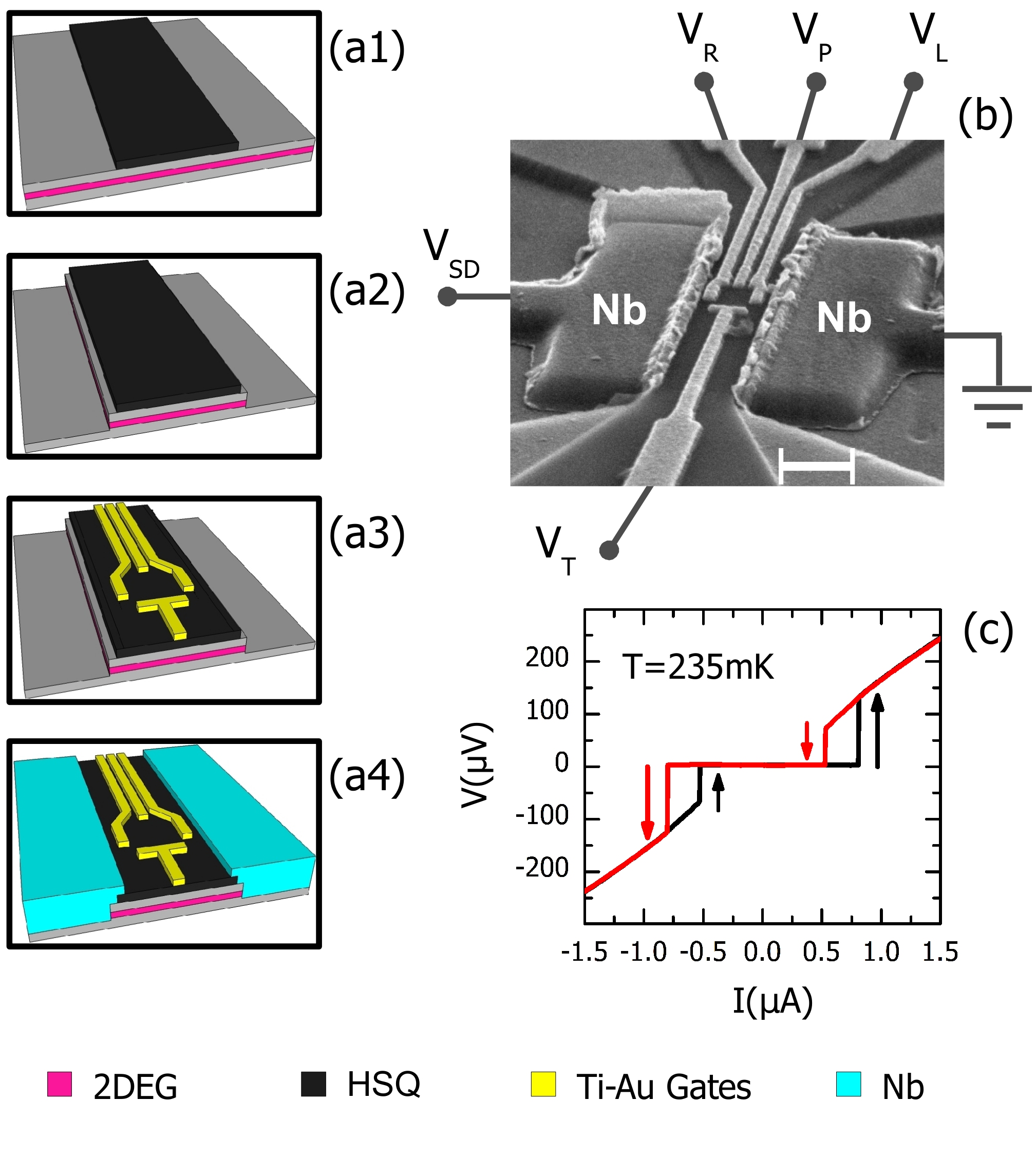}
\caption{\textbf{\textsf{Device layout.}}
(a1-4) Summary of the main nanofabrication steps and cross-sectional schematics of the hybrid superconductor-quantum dot device. (b) Scanning electron micrograph of the sample (scale bar is \SI{500}{n m}).
(c) Voltage-current characteristic at $T=\SI{235}{mK}$ of the open Nb-InGaAs-Nb junction (all gates set to ground). 
}
\end{figure}

When negative voltages $V_L$, $V_R$ and $V_T$ are applied to the electrostatic gates [see Fig. 1(b)], the underlying 2DEG regions are depleted, forming a QD. The voltage $V_P$ applied to the fourth gate allows to tune the charge state of the QD. Two point-contacts with tunable transparency couple the QD island to the nearby Nb-InGaAs (S-2DEG) junctions. The distance between the Nb-InGaAs interfaces and the electrostatic gates that define the QD is of the order of \SI{100}{nm}, below the induced coherence length $\xi_N$.  For superconductor-normal metal (SN) junctions with a normal region length approaching or smaller than $\xi_N$ the proximity-induced energy gap $\Delta^*$ is expected to increase, reaching $\Delta$ in the case of ideal NS interfaces in the short junction limit; a suppression of $\Delta^*$ should occur for contacts with finite interface transparency\cite{ref20}. It is known that the stability diagram of a QD directly connected to superconductors reveals the presence of a gap in the DOS, both for weak and strong  tunnel coupling to the leads\cite{ref4,ref9,ref21}. We will show in the following that our device can be easily tuned between the two regimes, and report our measurement of $\Delta^*$ in the 2DEG.

\begin{figure}[t!]\label{Fig2}
\includegraphics[width=85mm]{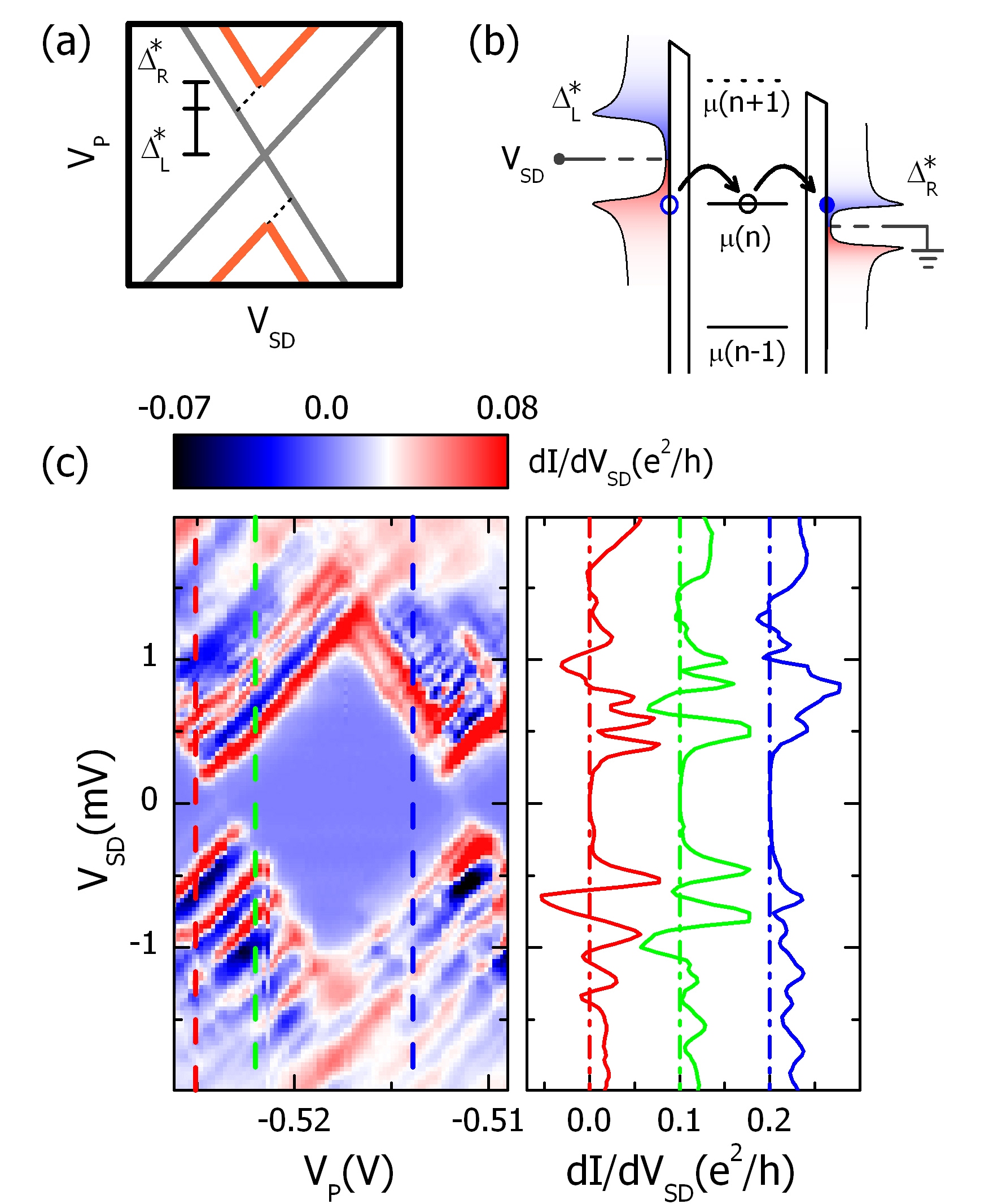}
\caption{\textbf{\textsf{Stability diagram for the closed QD.}}
(a) Sketch of the Coulomb diamond edges in the case of normal leads (gray lines) and in the presence of proximity-induced gaps $\Delta^*_L$ and $\Delta^*_R$. 
(b) energy diagram for single electron tunneling in the hybrid QD.
(c) Left panel: color plot of the differential conductance versus $V_{SD}$ and $V_P$ (stability diagram) in the case of weak coupling to the leads at $T=\SI{235}{mK}$. Right panel: $dI/dV_{SD}$ traces for three values of plunger gate voltage $V_P$ (indicated by dashed lines in the stability plot). The curves are horizontally shifted for clarity: dot-dashed lines indicate $dI/dV_{SD}=0$.
}
\end{figure}

Figure 2(a) shows measurements ($T=\SI{235}{m K}$) in a regime of weak QD-lead coupling: the QD differential conductance $dI/dV_{SD}$ is plotted in color scale as a function of the source-drain bias $V_{SD}$ and $V_P$ (stability plot). In the diamond-shaped regions of low conductance electron tunneling through the QD is forbidden by Coulomb repulsion, and the QD occupation number $n$ is constant. Diamond edges with positive (negative) slope correspond to the onset of single electron tunneling from the left (right) lead. A sketch of the stability plot is given in Fig. 2(b): in the case of normal contacts (gray lines) the diamond edges cross at $V_{SD}=0$, forming zero bias Coulomb blockade peaks at the charge-degeneracy points. The presence of gaps $\Delta^*_L$ and $\Delta^*_R$ in the DOS of the left and right lead, respectively, shifts the onset of the corresponding diamond edges [orange lines in Fig. 2(b)] to higher values of $|V_{SD}|$ [see Fig. 2(c)]. The measured values of $\Delta^*_L$ and $\Delta^*_R$ are $\Delta^*_L=(180\pm10)\SI{}{\micro eV}$ and $\Delta^*_R=(15\pm10)\SI{}{\micro eV}$. Such a large asymmetry could be linked to asymmetric tunnel coupling of the QD to the leads and to the consequent asymmetric depletion and weakening of the proximity effect in the adjacent 2DEG regions. Our finding that $\Delta^*_L\gg\Delta^*_R$ is also consistent with the fact that excited-state lines pertaining to the left lead are more intense and followed by regions of large negative $dI/dV_{SD}$\cite{ref9}.

\begin{figure}[t!]\label{Fig3}
\includegraphics[width=85mm]{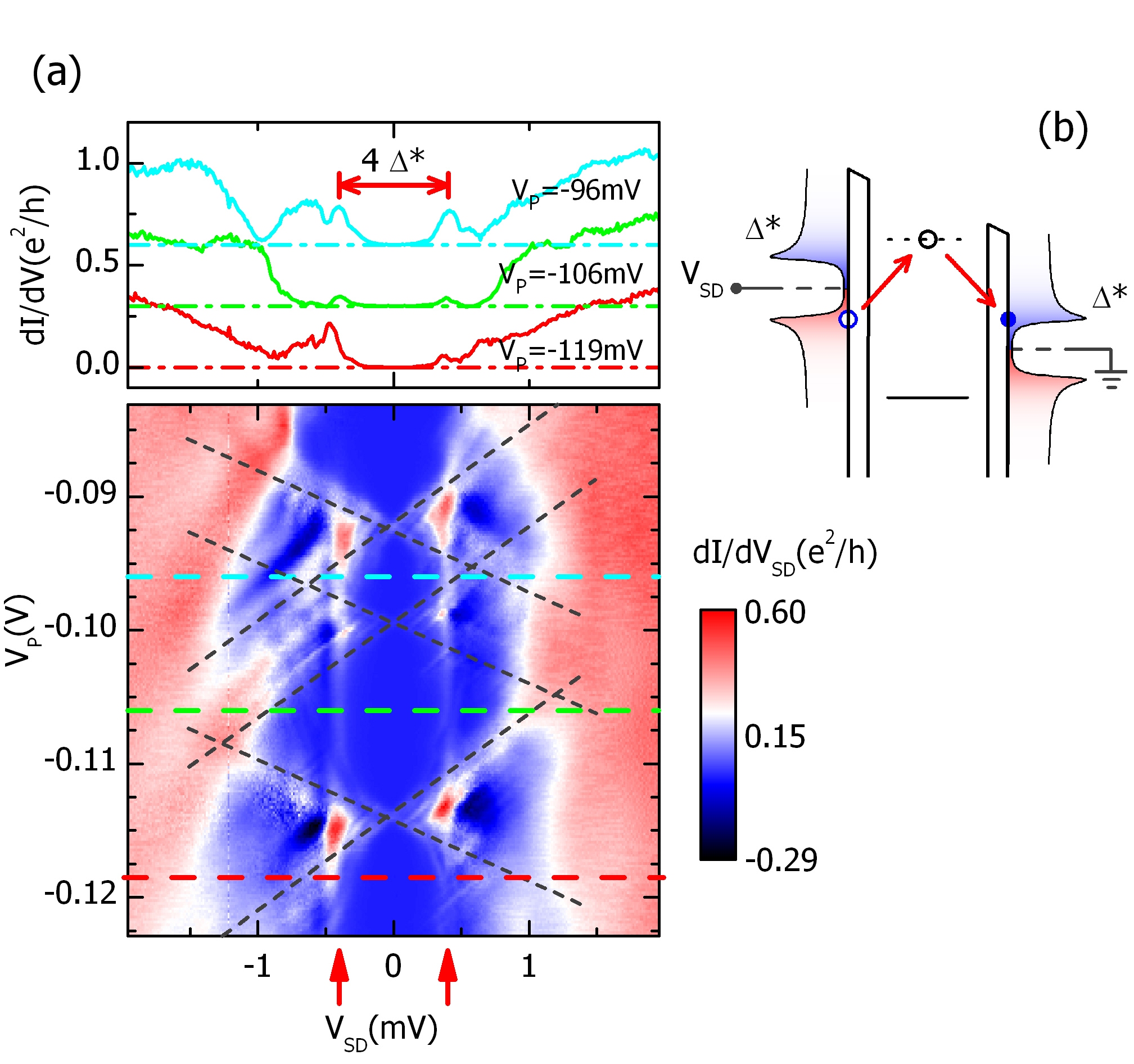}
\caption{\textbf{\textsf{Cotunneling spectroscopy.}}
(a) Stability diagram in the regime of stronger QD-lead coupling at $T=\SI{150}{mK}$: symmetric peaks (indicated by the red arrows) appear inside the diamonds at constant voltages $V_{SD}=\pm 2\Delta^*/e$ (independent of $V_P$), where the electron- and hole-like DOS peaks enhance the onset of elastic processes. One of such cotunneling processes is sketched in panel (b). Differential conductance profiles for three values of $V_P$ (indicated by dashed lines in the stability plot) are displayed in the upper panel. The curves are vertically shifted for clarity: dot-dashed lines in the upper panel indicate $dI/dV_{SD}=0$).}
\end{figure}

Our device architecture allows to easily tune the QD to a regime of stronger coupling to the leads by increasing the voltages $V_L$, $V_R$, $V_T$ of a few \SI{}{mV}. In this regime cotunneling processes lead to the appearance of finite conductance inside the Coulomb diamonds [see Fig. 3(a)]]. Symmetric peaks in $dI/dV_{SD}$ located at a constant voltage $V_{SD}=\SI{330}{\micro V}$, independent of the charge state of the QD, are due to elastic cotunneling processes, such as the one sketched Fig. 3(b). In the upper panel of Fig. 3(a) the differential conductance is plotted versus $V_{SD}$ for constant values of $V_P$ indicated by the dashed lines of corresponding color in the lower panel. In this case our data indicate that $\Delta^*_L=\Delta^*_R\equiv\Delta^*$: the onset of elastic cotunneling takes place at $|eV_{SD}|=2\Delta^*$, where the particle- and hole-like DOS peaks in the two leads are aligned. We thus find $\Delta^*=\SI{165}{\micro eV}$. This value is slightly smaller than the value of $\Delta^*_L$ found in the weak coupling regime (the difference being comparable with the uncertainty on $\Delta^*_L$). We note that the data shown in Fig. 2 and Fig. 3 were taken in different cooldowns of the device, and a fluctuation in $n_S$ could explain a variation in the strength of the proximity effect.

In conclusion, we demonstrated a hybrid QD device obtained by means of standard top-down nanofabrication, which combines a Nb-\iga{0.8}{0.2}-Nb planar junction with a lateral QD confined by electrostatic gates. Transport measurements allow to directly measure a proximity-induced gap in the 2DEG which results of the order of \SI{200}{\micro eV}. As these results are relevant to the development of non-dissipative single electron transistors, further work will be devoted to the study of Josephson coupling through our device.

The authors would like to thank P. Spathis for fruitful discussions and for help with the cryogenic setup. We acknowledge partial financial support from the E.U. Project HYSWITCH (grant No. FP6-517567), the MIUR-FIRB No. RBIN06JB4C, the INFM-CNR Seed project `Quantum dot refrigeration: accessing the \SI{}{\micro K} regime in solid-state nanosystems', and the NanoSciERA project `NanoFridge'.

%\bibliography{SQDS.bib}

\begin{thebibliography}{35}
\bibitem{ref18.6}
 \bibinfo{note}{The observed hysteresis is typical of S-N-S proximity junctions with high critical current density\cite{ref19}.}
\end{thebibliography}

\end{document}